\documentclass{aastex631}

\usepackage[utf8]{inputenc}
\usepackage{xcolor}
\usepackage{hyperref}
\usepackage{graphicx}
\usepackage{mathrsfs}

\def\vec#1{{\bf #1}}
\newcommand{\myurl}[2][red]{\href{#2}{\color{#1}{#2}}}%

\begin{document}

\received{}
\revised{}
\accepted{}
\submitjournal{ApJS}

\shorttitle{PHOEBE VIII: Multiple systems}
\shortauthors{Brož et al.}

\title{Physics Of Eclipsing Binaries. VIII. Multiple systems}

\correspondingauthor{Miroslav Brož}
\email{mira@sirrah.troja.mff.cuni.cz}

\author[0000-0003-2763-1411]{Miroslav Brož}
\affiliation{Charles University, Faculty of Mathematics and Physics, Institute of Astronomy, V Holešovičkách 2, CZ-18200 Praha 8}

\author[0000-0002-5442-8550]{Kyle E.~Conroy}
\affiliation{Villanova University, Dept.~of Astrophysics and Planetary Sciences, 800 E.\ Lancaster Ave, Villanova, PA 19085, USA}

\author[0000-0002-1913-0281]{Andrej Pr\v sa}
\affiliation{Villanova University, Dept.~of Astrophysics and Planetary Sciences, 800 E.\ Lancaster Ave, Villanova, PA 19085, USA}


\begin{abstract}
Multiple stellar systems are common especially among O and B stars.
In order to accurately describe their dynamics,
interactions among components must be accounted for.
In this work, we describe the new dynamical model in Phoebe,
which could be used just for this purpose.
The n-body model is based on the Reboundx numerical integrator 
and accounts for 
mutual perturbations,
oblateness,
relativistic effects, or
light-time effects.
The initial conditions
can be set up as hierarchical or two-pairs geometry.
For comparison purposes,
we also provide a simplified keplerian model.
Photometric computations work similarly as before,
with Roche distortions for pairs of components
(or for centres of mass, if hierarchical),
and all mutual eclipses.
If the time span of observations is long enough,
so that perturbations
(precession, resonances)
are manifested in eclipse timings or durations,
this allows to construct order-of-magnitude more precise models of stellar systems.
\textcolor{red}{This draft refers to a development version of Phoebe, available at \myurl{https://github.com/miroslavbroz/phoebe2/tree/interferometry}.
It is not yet included in the official Phoebe repository!}
\end{abstract}


\section{Introduction}

multiple stellar systems
common especially among O and B stars
\citep{Duchene_2013ARA&A..51..269D}
which are massive, compact, and with rotating components.
Generally, the problem of computing their interactions is too complex,
because each body becomes distorted by the gravitational potential,
determined by extended bodies,
which changes its internal structure,
and {\em vice versa\/}.

In the Roche approximation
\citep{Roche_1873,Kopal_1959cbs..book.....K,Wilson_1979ApJ...234.1054W}
the potential is point mass.
The rotation of bodies --if synchronous--
is accounted for as the centrifugal term.
For asynchronous or misaligned components,
the potential function is more complicated
\cite{Horvat_2018ApJS..237...26H}.
The true surfaces of stars are assumed to align with the respective iso-surfaces.

This description is used in Phoebe
\citep{Prsa_2016ApJS..227...29P,Horvat_2018ApJS..237...26H,Jones_2020ApJS..247...63J,Conroy_2020ApJS..250...34C},
a comprehensive model for binaries.
Even though its photometric algorithm is state-of-the-art,
its underlying dynamical model has not been extended for $n$ bodies.
An alternative approach was presented by
\cite{Broz_2017ApJS..230...19B,Broz_2021A&A...653A..56B,Broz_2022A&A...657A..76B,Broz_2022A&A...666A..24B,Broz_2023A&A...676A..60B}.
However, their n-body model is combined
with the original photometric algorithm of
\cite{Wilson_1971ApJ...166..605W},
which is, unfortunately, inferior to Phoebe.

Our motivation is to combine precise dynamics
with precise photometric computations,
so that Phoebe can be used in a wider range of applications,
from multiple massive stars
to interacting exoplanet systems
(like TOI-178; \citealt{Leleu_2021A&A...649A..26L,Delrez_2023A&A...678A.200D}).
Hereinafter, we describe the new dynamical model in Phoebe,
which is built on the Reboundx numerical integrator
\citep{Rein_2012A&A...537A.128R,Tamayo_2020MNRAS.491.2885T}.
For comparison purposes,
we also provide a simplified keplerian model.
We present both methods in Sec.~\ref{methods},
together with examples in Sec.~\ref{examples}.


\section{Methods}\label{methods}

\subsection{Keplerian model}\label{complex}

Even though a keplerian model \citep{Kepler_1619ikhm.book.....K}
is a solution for just two bodies,
it is often used for more than two,
provided the system is dominated by the central body
(e.g., the solar system)
or is hierarchical.

\paragraph{Hierarchy}

The first step is thus to build a hierarchy in Phoebe,
which is a description, how bodies are connected to orbits in
binaries, triples, quadruples, \dots\
For example, a triple system looks like

\begin{verbatim}
    orbit:orbit2
    
        orbit:orbit1
        
            star:starA
            star:starB
        
        star:starC
\end{verbatim}

Of course, the respective numbers of ${\rm orbits} = {\rm bodies} - 1$.
Hereinafter, we prefer indexing from $1$ (in the text),
even though it is indexed from $0$ (in Python).


\paragraph{Jacobi coordinates}

The second step is to define the Jacobi coordinates;
each body's coordinates (and velocities)
are referred to with respect to the centre of mass of all the preceding bodies.
This approach assures that one is as close as possible
to the original two-body solution.
Orbital elements are also defined in these coordinates.

In our model, we thus compute the sum of masses
$$m_{\rm sum} = m_1 + m_2\,,$$
the mean motion
$$n = \sqrt{Gm_{\rm sum}\over a^3}\,,$$
the period
$$P = {2\pi\over n}\,,$$
accounting for a possibility of the non-zero period rate
$$P' = P + \dot P\,(t-t_0)\,,$$
the corresponding semimajor axis
$$a' = a\left({P'\over P}\right)^2\,,$$
the mean anomaly (quadratic if $\dot P\ne 0$)
$$M' = M + n(t-t_0) - {n\dot P\over 2P}(t-t_0)^2\,,$$
the pericentre precession rate
$$\omega' = \omega + \dot\omega\,(t-t_0)\,,$$
and so on for next orbits.
The set of elements for each body is denoted as
\begin{equation}
\mathscr{E}_i \equiv \{a'_i, e_i, I_i, \Omega_i, \omega_i', M_i'\}\,.
\end{equation}

We use the standard subroutines from SWIFT \citep{Levison_1994Icar..108...18L},
rewritten from Fortran to Python,
to compute coordinates (and velocities)
\begin{equation}
\vec r_{{\rm j},i}, \vec v_{{\rm j},i} = \verb|orbel.orbel_el2xv|\left( {\textstyle\sum_{j=1}^i} Gm_j, \mathscr{E}_i \right) \quad {\rm for} \quad i = 2 .. n\,, \label{rj_vj}
\end{equation}
where
$\vec r_{{\rm j},1}, \vec v_{{\rm j},1} = 0$.


\paragraph{Barycentric coordinates}

Similarly, a conversion to the barycentric coordinates is done as
\begin{equation}
\vec r_{{\rm b},:}, \vec v_{{\rm b},:} = \verb|coord.coord_j2b|(Gm_:, \vec r_{{\rm j},:}, \vec v_{{\rm j},:})\,. \label{rb_vb}
\end{equation}

The $\gamma$ velocity is radial, which corresponds to the $\hat z$ direction in Phoebe
\begin{eqnarray}
\vec r_{{\rm b},:}' &=& \vec r_{{\rm b},:} - \gamma(t-t_0)\hat z\,, \label{rb_}\\
\vec v_{{\rm b},:}' &=& \vec v_{{\rm b},:} - \gamma\hat z\,. \label{vb_}
\end{eqnarray}

Moreover, we modify the signs of $x$, $y$,
due to the coordinate convention in Phoebe
\begin{eqnarray}
\vec r_{{\rm b},:}'' &=& \vec r_{{\rm b},:}' \times (-1, -1, +1)\,, \label{rb__}\\
\vec v_{{\rm b},:}'' &=& \vec v_{{\rm b},:}' \times (-1, -1, +1)\,. \label{vb__}
\end{eqnarray}


\paragraph{Euler angles}

The Euler angles are needed for proper orientations of non-spherical bodies.
Using the Kepler equation, we obtain the eccentric anomaly
\begin{equation}
E = \verb|orbel.orbel_ehie|(e, M)\,,
\end{equation}
then the true anomaly
\begin{equation}
\theta = 2\arctan\left(\sqrt{1+e\over 1-e}\tan{E\over 2}\right)\,,
\end{equation}
and the Euler angles as
\begin{eqnarray}
{\rm euler}_{1,i} &=& \cases{
\theta_i + \omega_i & for $i = 1$\,,\cr
\theta_i + \omega_i + \pi & for $i = 2..n$\,,
} \\
{\rm euler}_{2,i} &=& M_i\,, \\
{\rm euler}_{3,i} &=& I_i\,,
\end{eqnarray}
respectively.


\paragraph{Constraints}

Apart from the existing constraints for binaries,
additional constraints are necessary for triples,
between the masses and the mass ratio
\begin{equation}
m_3 = q_2(m_1+m_2)\,, \label{m_3}
\end{equation}
between the periods and semimajor axes
\begin{equation}
a_2 = \left[{a_1^3\over P_1^2} P_2^2 (1+q_2)\right]^{1\over 3}\,, \label{a_2}
\end{equation}
which prevents setting up contradicting parameter values.


\subsection{N-body model}\label{simplified}

An n-body model is a numerical solution for $n$ bodies,
which is much more general.

\paragraph{Geometry}

To set up initial conditions for a numerical integration,
we use the same sets of elements
(without $'$ quantities though)
\begin{equation}
\mathscr{E}_i \equiv \{a_i, e_i, I_i, \Omega_i, \omega_i, M_i\}\,;
\end{equation}
they are considered as osculating,
defined only for $t = t_0$.
Internally, we use the units of au, d, rad, $M_\odot$, and $G = 1$.

For a hierarchical geometry,
Eq.~(\ref{rj_vj}),
Eq.~(\ref{rb_vb}), and
Eqs.~(\ref{rb__}-\ref{vb__})
are used to obtain the initial coordinates and velocities.

For a two-pairs geometry,
a series of keplerian models is used as follows,
\begin{eqnarray}
\vec r_{2,1}, \vec v_{2,1} &=& \verb|orbel.orbel_el2xv|\left({\textstyle\sum_{j=1}^2} Gm_j, \mathscr{E}_0\right)\,, \\
\vec r_{4,3}, \vec v_{4,3} &=& \verb|orbel.orbel_el2xv|\left({\textstyle\sum_{j=3}^4} Gm_j, \mathscr{E}_1\right)\,, \\
\vec r_{34,12}, \vec v_{34,12} &=& \verb|orbel.orbel_el2xv|\left({\textstyle\sum_{j=1}^4} Gm_j, \mathscr{E}_2\right)\,.
\end{eqnarray}

The corresponding `heliocentric' coordinates (and velocities) are
\begin{eqnarray}
\vec r_{{\rm h},1} &=& \vec 0\,, \nonumber\\
\vec r_{{\rm h},2} &=& \vec r_{2,1}\,, \nonumber\\
\vec r_{{\rm h},3} &=& \vec r_{12,1} + \vec r_{34,12} - \vec r_{34,3}\,, \nonumber\\
\vec r_{{\rm h},4} &=& \vec r_{{\rm h},2} + \vec r_{4,3}\,, \nonumber
\end{eqnarray}
which are converted to the Jacobi coordinates
\begin{equation}
\vec r_{{\rm j},:}, \vec v_{{\rm j},:} = \verb|coord.coord_h2j|(Gm_:, \vec r_{{\rm h},:}, \vec v_{{\rm h},:})\,,
\end{equation}
and eventually to the barycentric coordinates
\begin{equation}
\vec r_{{\rm b},:}, \vec v_{{\rm b},:} = \verb|coord.coord_j2b|(Gm_:, \vec r_{{\rm j},:}, \vec v_{{\rm j},:})\,.
\end{equation}


\paragraph{Roche parameters}

Since orbital elements are no longer constants,
we have to evaluate the Roche parameters
for each time.
We thus implemented an inverse geometry,
to recover the sets of elements
$$\mathscr{E}_i(t) = \{a_i(t), e_i(t), I_i(t), \Omega_i(t), \omega_i(t), M_i(t)\}\,.$$

The respective Roche parameters are defined as the relative separation and the synchronicity
\begin{eqnarray}
{\rm roche}_{1,i} &=& r_i/a_i\,, \\
{\rm roche}_{2,i} &=& P_i/P_\star\,,
\end{eqnarray}
where the mean motion
$n_i = \sqrt{\sum_{j=1}^i Gm_j/a_i^3}$,
and the period
$P_i = 2\pi/n_i$.


\paragraph{Constraints}

For hierachical systems,
Eq.~(\ref{m_3}) and
Eq.~(\ref{a_2})
are applied sequentially.

For two pairs, the hierarchy is different,
and one constraint must have been added,
\begin{equation}
q_3 = {m_3+m_4\over m_1+m_2}\,,
\end{equation}
in order to constrain
\begin{equation}
a_3 = \left[{a_1^3\over P_1^2} P_3^2 (1+q_3)\right]^{1\over 3}\,.
\end{equation}


\paragraph{Perturbations}

We use the Rebound or Reboundx numerical integrator
\citep{Rein_2012A&A...537A.128R,Tamayo_2020MNRAS.491.2885T}
to solve the following equation of motion
\begin{equation}
\ddot\vec r_i = -\sum_{j\ne i} {Gm_j\over r_{j,i}^3} \vec r_{j,i} + \vec f_{\rm oblat} + \vec f_{\rm ppn}\,,
\end{equation}
where the first term corresponds to n-body perturbations,
the second term to oblateness,
and the third term to relativistic effects.

The n-body perturbations self-consistently include
precession of $\Omega$, $\omega$,
variation,
evection,
Kozai cycles,
close encounters,
hyperbolic trajectories,
mean-motion resonances,
secular resonances,
three-body resonances, or
chaotic diffusion due to overlapping resonances.

The time step $\Delta t$ is not fixed,
but adaptively adjusted,
according to the relative precision $\epsilon$.
In most applications,
$\epsilon = 10^{-9}$ should be sufficient,
but it is advisable to compute a convergence test.


\paragraph{Oblateness}

The oblateness computation in Reboundx was rewritten,
to allow for a general (i.e., not $\hat z$) orientation of the spin axis.
This was necessary, because orbits in Phoebe can be in any plane
and bodies can be misaligned.

Let us take any two bodies ($i$, $j$),
where the 1st acts on the 2nd.
Its spin-axis unitvector
\begin{equation}
\hat s = R_z(\Omega_\star) \times R_x(-i_\star) \times \hat z\,,
\end{equation}
determines the new basis
\begin{eqnarray}
\hat w &=& \hat s\,, \\
\hat u &=& (-s_x, s_y, 0)\,, \\
\hat v &=& -\hat u\times\hat w\,.
\end{eqnarray}
The new coordinates are
\begin{eqnarray}
u &=& \hat u\cdot\vec r_{j,i}\,, \nonumber\\
v &=& \hat v\cdot\vec r_{j,i}\,, \nonumber\\
w &=& \hat w\cdot\vec r_{j,i}\,. \nonumber
\end{eqnarray}
In these coordinates, the acceleration is simple,
\begin{eqnarray}
f_1 &=& {\textstyle{3\over 2}} Gm_i J_{2,i} R_i^2/r_{j,i}^5\,, \\
f_2 &=& 5\cos\vartheta^2 - 1\,, \\
f_3 &=& f_2 - 2\,,
\end{eqnarray}
where
$J_{2,i} = -C_{20,i}$ is the dimension-less parameter,
$R_i$ the reference radius,
$\cos\vartheta = w/|\vec r_{j,i}|$ the inclination with respect to the equator,
and
\begin{equation}
\vec f_{uvw} = (f_1f_2u, f_1f_2v, f_1f_3w)\,.
\end{equation}

Since the old basis (') in new coordinates is
\begin{eqnarray}
\hat x' &=& (u_x, v_x, w_x)\,, \nonumber\\
\hat y' &=& (u_y, v_y, w_y)\,, \nonumber\\
\hat z' &=& (u_z, v_z, w_z)\,, \nonumber
\end{eqnarray}
the acceleration in old coordinates
\begin{equation}
\vec f_{xyz} = (\hat x'\cdot\vec f_{uvw}, \hat y'\cdot\vec f_{uvw}, \hat z'\cdot\vec f_{uvw})\,.
\end{equation}
Again, $\sum_{j\ne i}$, $\forall i$ must be computed.
In particular, the oblateness induces additional precession of $\Omega$, $\omega$.
An approximation of no back-reaction, zero torque,
and no evolution of the spin axis is used.


\paragraph{Relativistic effects}

The parametrized post-newtonian (PPN) approximation
is used to account for relativistic effects in Reboundx.
For any two bodies ($i$, $j$),
the acceleration is \citep{Newhall_1983A&A...125..150N}
\begin{eqnarray}
\vec f_{\rm ppn} &=& \sum_{j\ne i} {\mu_j (\vec r_j-\vec r_i)\over r_{ij}^3}
\Biggl\{- {2(\beta+\gamma)\over c^2} \sum_{k\ne i} {\mu_k\over r_{ik}}
- {(2\beta - 1)\over c^2} \sum_{k\ne j} {\mu_k\over r_{jk}}
+ \gamma \left(v_i\over c\right)^2
+ (1+\gamma) \left(v_j\over c\right)^2
- {2(1+\gamma)\over c^2} \dot\vec r_i\cdot\dot\vec r_j \,- \nonumber\\
&-& {3\over 2c^2} \left[{(\vec r_i-\vec r_j) \cdot \dot\vec r_j \over r_{ij}}\right]^2
+ {1\over 2c^2}(\vec r_j - \vec r_i) \cdot \ddot\vec r_j \Biggr\}
+ {1\over c^2} \sum_{j\ne i} {\mu_j\over r_{ij}^3} \left\{\left[\vec r_i-\vec r_j\right] \cdot \left[(2+2\gamma)\dot\vec r_i - (1+2\gamma)\dot\vec r_j\right]\right\} (\dot\vec r_i-\dot\vec r_j) \,+ \nonumber\\
&+& {3+4\gamma\over 2c^2} \sum_{j\ne i} {\mu_j\ddot\vec r_j\over r_{ij}}\,, \label{ppn}
\end{eqnarray}
where
$c$ is the speed of light,
$\beta \equiv v/c$,
$\gamma \equiv 1/\sqrt{1-\beta^2}$.,
Again, $\sum_{j\ne i}$, $\forall i$ must be computed.
The relativistic effects primarily induce precession of $\omega$.


\paragraph{Light-time effects}

The light-time effects
\citep{Romer_1677,Rappaport_2013ApJ...768...33R}
are important especially for extended systems,
where the time must be referred to with respect to the barycentre.
For each body (separately), the proper time is estimated as
\begin{equation}
t_{\rm proper} - {z(t_{\rm proper})\over c}\,{{\rm au}\over {\rm day}} - t_{\rm obs} = 0\,.
\end{equation}
Since the $z$ coordinate is time-dependent,
$t_{\rm proper}$ is solved for by the Newton method.

A barycentric correction w.r.t. the Earth must be applied separately
(it is not accounted for in Phoebe).


\subsection{Implementation notes}

The keplerian dynamics in Phoebe was rewritten,
to assure a 1:1 correspondence to n-body dynamics.
In the course of rewriting,
a $10^{-8}$ round-off error of $\pi$ has been corrected (in constraints).
We introduced a new geometry layer,
which is used to convert orbital elements to $xyz$ coordinates,
and the corresponding "xyz" integrator.
This is a useful generalisation,
allowing for various definitions of orbital elements.

A triple system in Phoebe is set up by
\verb|b = phoebe.default_triple()|.
In order to compute it with the n-body model,
one has to use
\verb|b.add_compute(dynamics_method='rebound')|.
Alternatively,
if the geometry layer should be used,
\verb|b.add_compute(dynamics_method='xyz', geometry='hierarchical')|.
It is always necessary to assure passing the \verb|geometry| keyword ('hierarchical', 'twopairs')
in agreement with the hierarchy.

The integrators in Rebound can be selected by passing the
\verb|integrator| keyword ('ias15', 'whfast', 'sei', 'leapfrog', 'hermes').
The default value of the initial step size is
$\Delta t = 0.01\,{\rm d}$,
and the relative precision
$\epsilon = 10^{-9}$.


\section{Examples} \label{examples}

\subsection{Comparison of keplerian vs. n-body models}

\begin{figure}
\centering
\includegraphics[width=6cm]{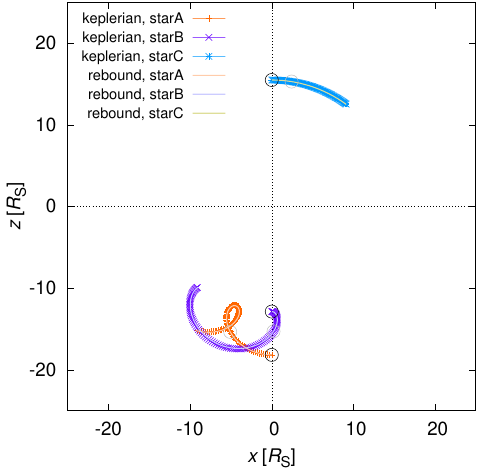}
\includegraphics[width=6cm]{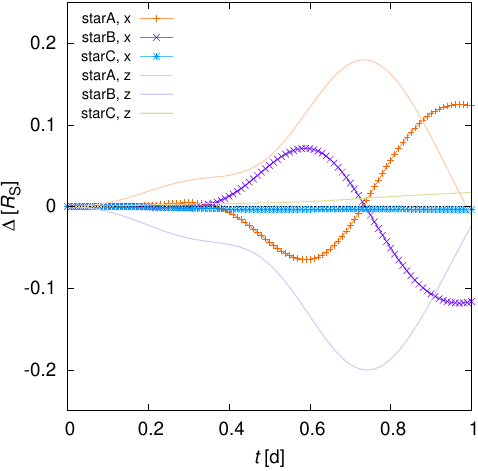}
\caption{
Triple system example,
with a comparison of keplerian (points) and n-body (lines) models.
The default triple is used
(i.e.,
$P_1 = 1\,{\rm d}$,
$P_2 = 10\,{\rm d}$,
$q_1 = 1$,
$q_2 = 1$,
$a_1 = 5.3\,R_\odot$,
$a_2 = 30.994588\,R_\odot$,
$m_1 \doteq 0.998813\,M_\odot$,
$m_2 \doteq 0.998813\,M_\odot$,
$m_3 \doteq 1.997626\,M_\odot$);
with a hierarchical geometry.
The initial conditions were exactly the same.
However, this system is 'too' compact,
and the zero precession ($\dot\omega = 0$) in the keplerian model
leads to substantial differences (cf.~$\Delta$).
}
\label{test_xyz_3body}
\end{figure}

A comparison of the keplerian and n-body models confirms
that the initial conditions are exactly the same
(see Fig.~\ref{test_xyz_3body}).
However, the temporal evolution is not the same,
because the default triple is too compact.
The precession rates ($\dot\omega_1$, $\dot\omega_2$)
are substantial on the orbital time scale.
To this point, we verified a that the more distant tertiary,
the better the correspondence between the models.

\subsection{Two pairs}

\begin{figure}
\centering
\includegraphics[width=9cm]{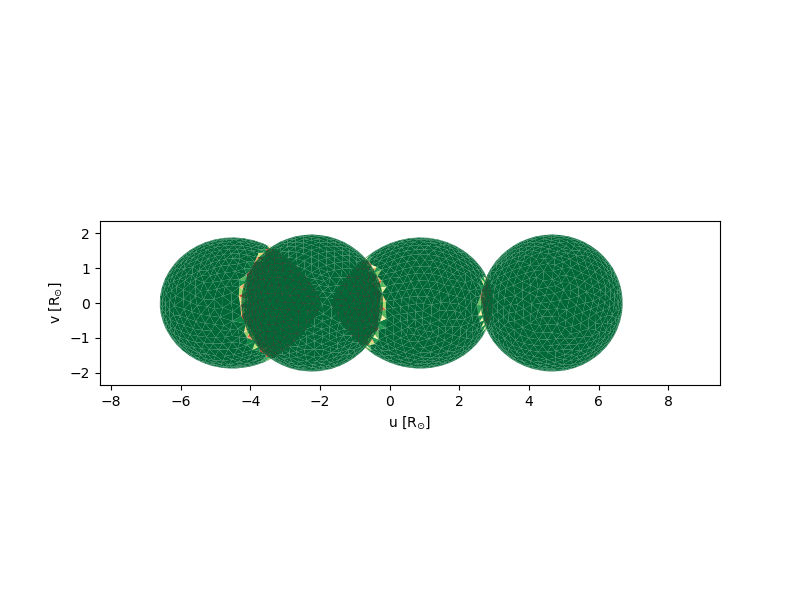}
\includegraphics[width=6cm]{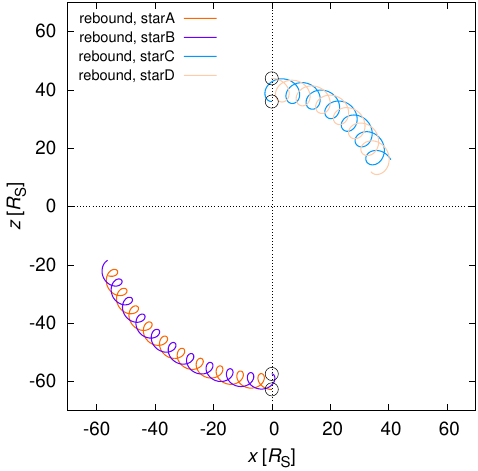}
\caption{
Two-pairs system example,
with modified parameters
($P_1 = 1\,{\rm d}$,
$P_2 = 1.5\,{\rm d}$,
$P_3 = 100\,{\rm d}$,
$q_1 = 1$,
$q_2 = 1$,
$q_3 = 1$,
$a_1 = 5.3\,{\rm au}$,
$a_2 = 7.95\,{\rm au}$,
$a_3 \doteq 154.972940\,{\rm au}$,
$m_1 \doteq 0.998813\,M_\odot$,
$m_2 \doteq 0.998813\,M_\odot$,
$m_3 \doteq 1.498219\,M_\odot$,
$m_4 \doteq 1.498219\,M_\odot$).
One of the pairs is close,
so the Roche lobes are clearly visible;
the other pair less so.
}
\label{test_xyz_twopairs}
\end{figure}

For two pairs,
only the n-body model was used,
with the corresponding geometry.
Its orbital evolution, shown in Fig.~\ref{test_xyz_twopairs}, is as expected.
Since the mutual period ($P_3$) was set long,
each of the three orbits would be close to keplerian.

\subsection{Eclipses}

\begin{figure}
\centering
\includegraphics[width=8cm]{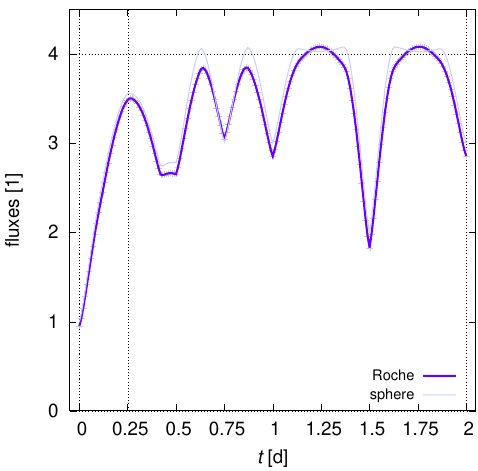}
\caption{
Light curve of the two-pairs system,
with multiply eclipsing components.
A comparison of two distortion methods is shown,
Roche (\color{blue}blue\color{black})
vs. spherical approximation (\color{gray}gray\color{black}).
The time $0.25\,{\rm d}$ is shown in Fig.~\ref{test_xyz_twopairs}.
}
\label{test_xyz_twopairs_LC}
\end{figure}

For the two pairs,
we also computed a light curve
(Fig.~\ref{test_xyz_twopairs_LC}).
It shows a complex behaviour due to
eclipses of each pair,
as well as multiple eclipses of components due to the other pair.
These photometric computations are relatively slow, in particular
the radiosity problem for $n$ bodies
integrals over triangles and
fractions of triangles,
which are visible,
since the algorithm's complexity is ${\cal O}(N^2)$,
where $N$ denotes the number of triangles.

A comparison of two distortion methods,
Roche vs. spherical approximation,
demonstrates how substantial could be the differences
(Fig.~\ref{test_xyz_twopairs_LC}).
When some of the components fill the Roche lobe,
the relative flux at distinct phases differs by ${\approx}\,0.2$.
Of course, this is orders of magnitude more than
a common precision of space-based photometric measurements by
MOST \citep{Walker_2003PASP..115.1023W},
Corot \citep{Auvergne_2009A&A...506..411A},
Kepler \citep{Borucki_2011ApJ...736...19B}, or
TESS \citep{Ricker_2015JATIS...1a4003R}.

\subsection{Roche distortion}

\begin{figure}
\centering
\includegraphics[width=9cm]{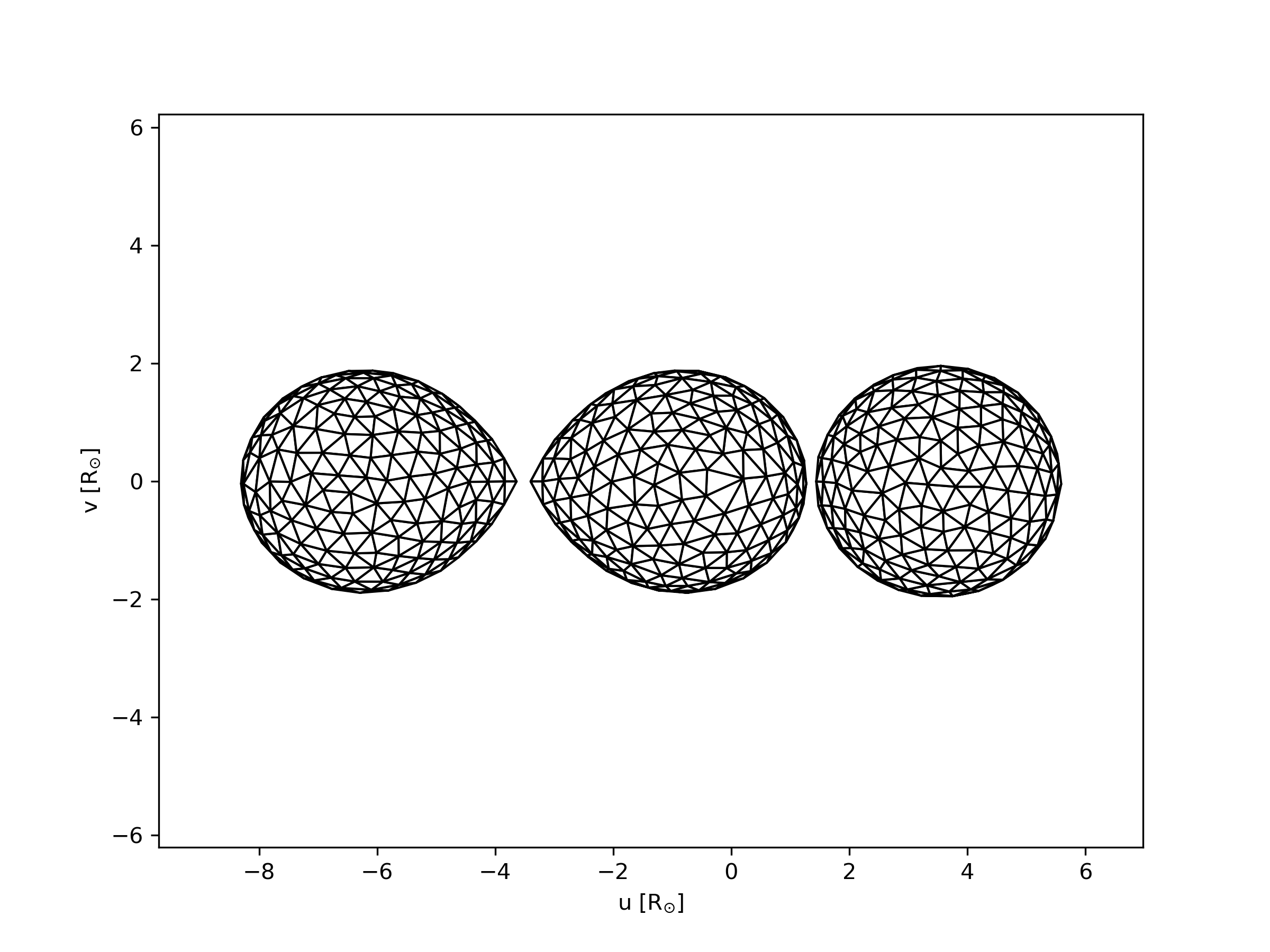}
\caption{
Compact system example,
with an artificially close tertiary.
A centre-of-mass approximation was used to compute its distortion;
in fact, the right component should be similar to the left component
and the middle component should be bi-lobed.
}
\label{test_phoebe18_compact}
\end{figure}

On the other hand, to demonstrate a limitation of our model,
we set up an artificially compact triple
(Fig.~\ref{test_phoebe18_compact}).
The Roche distortion is computed pair by pair,
which implies that
(i)~the two components are not influenced of the 3rd component;
(ii)~the 3rd component is influenced by their centre of mass.
In fact, all $n$ components contribute to the total potential,
for which all the iso-surfaces should be computed.
Nevertheless, the current approach
\citep{Horvat_2018ApJS..237...26H}
is considered as sufficient in most situations.

\subsection{Oblateness}

\begin{figure}
\centering
\includegraphics[width=6cm]{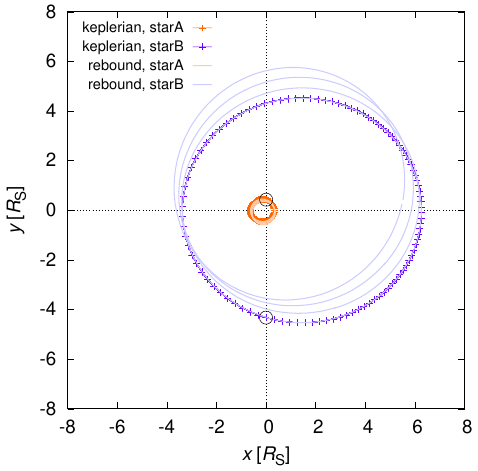}
\includegraphics[width=6cm]{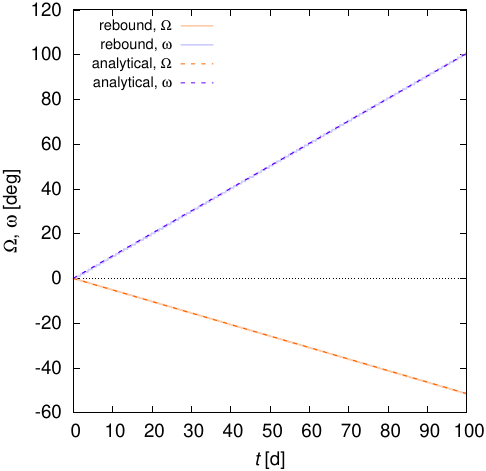}
\caption{
Oblateness dynamics example,
computed for an eccentric and inclined binary
($P_1 = 1\,{\rm d}$,
$e_1 = 0.3$,
$i_1 = 10^\circ$
$\Omega_1 = 0^\circ$,
$\omega_1 = 0^\circ$,
$q_1 = 0.1$,
$m_1 = 1.816023\,M_\odot$,
$m_2 = 0.181602\,M_\odot$,
$R_1 = 1.5\,R_\odot$,
$R_2 = 0.5\,R_\odot$).
The oblateness of the primary was
$J_2 = 0.01$.
The pitch angle of the primary,
$i_\star = -10^\circ$,
so that the spin~$\hat s$ is aligned with the $\hat z$-axis.
Otherwise, serious projection effects occur,
because the orbital elements ($i$, $\Omega$, $\omega$)
are defined with respect to the sky plane.
The precession rates ($\dot\Omega$, $\dot\omega$) are in agreement
with the analytical predictions
(dashed; e.g., \citealt{Oplistilova_2023A&A...672A..31O}).
}
\label{test_xyz_j2_Oplistilova}
\end{figure}

The oblateness effects were computed for an oblate primary
and a point-like secondary,
orbiting on an eccentric and inclined orbit
(Fig.~\ref{test_xyz_j2_Oplistilova}).
The precession rates of $\omega$, $\Omega$
were verified against the analytical relations
(e.g., \citealt{Oplistilova_2023A&A...672A..31O})
\begin{equation}
\dot\omega_1 = +3n_1J_2 \left({R_1\over a_1}\right)^2 {5\cos^2\vartheta-1\over 4\eta_1^4}\,,
\end{equation}
\begin{equation}
\dot\Omega_1 = -{3\over 2}n_1J_2 \left({R_1\over a_1}\right)^2 {\cos\vartheta\over\eta_1^4}\,,
\end{equation}
where
$n_1 = \sqrt{G(m_1+m_2)/a_1^3}$
$\eta_1 \equiv \sqrt{1-e_1^2}$,
and $\vartheta$ is the inclination with respect to the primary equator.
The agreement is excellent (Fig.~\ref{test_xyz_j2_Oplistilova});
the n-body model exhibits only tiny oscillations on the orbital time scale.

It is important to note that orbital elements in Phoebe are strictly defined
with respect to the sky plane. This implies serious projection effects
if the oblate body spin $\hat s$ is not aligned with the $\hat z$ axis.
For example, if $\hat s = \hat x$, $\vartheta = 10^\circ$,
$\Omega$ will only vary in a limited interval of $\pm 10^\circ$.
By default, spins $\hat s$ are perpendicular to the orbital planes.
This implies zero nodal precession, as the node is undefined.

\subsection{Relativistic effects}

\begin{figure}
\centering
\includegraphics[width=6cm]{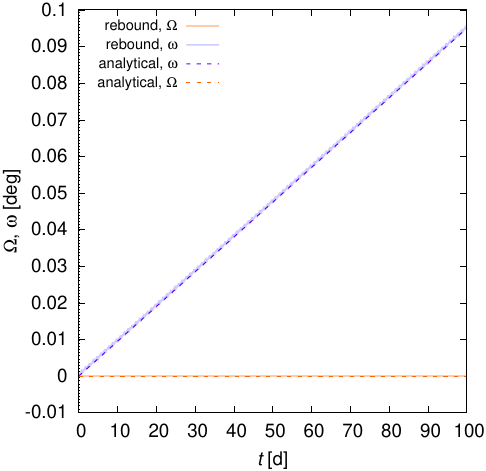}
\caption{
Relativistic effects example,
computed for the eccentric and inclined binary.
The precession rate ($\dot\omega$) is in agreement
with the analytical prediction (dashed).
Again, projection effects occur,
if $\hat s\ne\hat z$.
}
\label{test_xyz_gr_ecc}
\end{figure}

Likewise, the relativistic effects were computed for
the eccentric and inclined binary
(Fig.~\ref{test_xyz_gr_ecc}).
The precession rate of $\omega$
was verified against the analytical relation
\citep{Einstein_1916AnP...354..769E}
\begin{equation}
\dot\omega = {6\pi G\over c^2}{m_1+m_2\over a_1P_1(1-e^2)}\,.
\end{equation}
The agreement is again excellent (Fig.~\ref{test_xyz_gr_ecc}).
Only if all three contributions to precession are accounted for,
it is possible to obtain unbiased values of stellar parameters.

\subsection{Light-time effects}

\begin{figure}
\centering
\includegraphics[width=6cm]{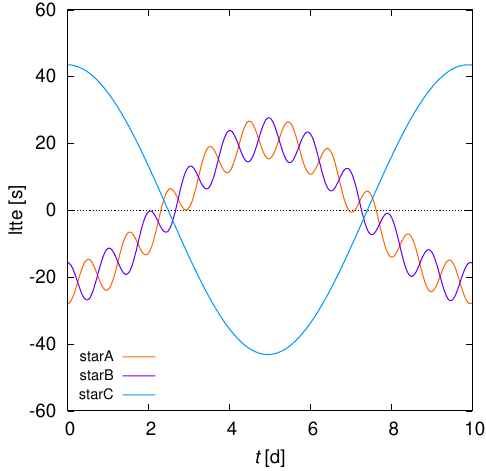}
\caption{
Light-time effects example,
computed for the default triple.
The respective delay (in seconds)
with respect to the barycentre
is plotted separately for the three components.
}
\label{test_xyz_ltte}
\end{figure}

Eventually, the light-time effects were evaluated,
in order to verify that the proper time is correctly estimated.
If \verb|conf.devel_on()| is used,
the the proper time minus the observed time is output
(see Fig.~\ref{test_xyz_ltte}).
Both the amplitudes, signs, and resulting shifts of coordinates
correspond to the physical dimension of the system.

All examples from Sec.~\ref{examples} are available.%
\footnote{\url{https://sirrah.troja.mff.cuni.cz/~mira/tmp/phoebe2/}}

\section{Conclusions}

We have described,
how multiple systems were incorporated in Phoebe,
with a number of examples.
While the dynamics itself is fast to compute,
corresponding photometric computations are relatively slow, in particular
the radiosity problem, 
integrals over triangles and
fractions of triangles,
which are visible,
since the algorithm's complexity is ${\cal O}(N^2)$. 
Nevertheless, it is still useful for precise modeling, whenever
the Roche distortion,
multiple eclipses,
dynamical effects, or
light-time effects
play a non-negligible role.
On top of that,
Phoebe can be used for benchmarking of other,
less precise, approximative models.










\begin{acknowledgements}
This work has been supported by the Czech Science Foundation through
grant 25-16507S (M. Brož).
\end{acknowledgements}

\bibliographystyle{apj}
\bibliography{paper2}

\end{document}